\documentclass[twocolumn, aps, nofootinbib]{revtex4}
\pdfoutput=1

\newcommand{\pprime}{{\prime\prime}}
\newcommand{\mc}[1]{{\mathcal{#1}}}
\newcommand{\vphi}{\varphi}
\usepackage{graphicx}
\usepackage{multirow}
\begin{document}

\title{Surprising phenomena in a rich new class of inflationary models}
\author{Pascal M.~Vaudrevange}\affiliation{Department of Physics \& CERCA, Case Western Reserve University, 10900 Euclid Ave, Cleveland, OH 44106}
\author{Dmitry I.~Podolsky}\affiliation{Department of Physics \& CERCA, Case Western Reserve University, 10900 Euclid Ave, Cleveland, OH 44106}
\author{Glenn D.~Starkman}\affiliation{Department of Physics, CERCA \& ISO, Case Western Reserve University, 10900 Euclid Ave, Cleveland, OH 44106}
\date{\today}

\begin{abstract}
  We report on a new class of fast-roll inflationary models.  In a
  huge part of its parameter space, inflationary perturbations exhibit
  quite unusual phenomena such as scalar and tensor modes freezing out
  at widely different times, as well as scalar modes reentering the
  horizon during inflation. 

  One specific point in parameter space is characterized by
  extraordinary behavior of the scalar perturbations. Freeze-out of
  scalar perturbations as well as particle production at horizon
  crossing are absent. Also the behavior of the perturbations around
  this quasi-de Sitter background is dual to a quantum field theory in
  flat space-time. Finally, the form of the primordial power spectrum
  is determined by the interaction between different modes of scalar
  perturbations.
\end{abstract}

\maketitle

\section{Introduction}
Inflation has become the standard paradigm for the description of the
early universe, solving the monopole, horizon and flatness problems as
well as providing for a mechanism to seed structure in the
universe. Various explicit models for this period of exponential
expansion, most relying on the dynamics of real scalar ``inflaton''
fields, mostly agree with current observations of perturbations of an
isotropic background, such as cosmic microwave background (CMB)
fluctuations and large scale structure (LSS) (although see
\cite{Copi:2008hw} and references therein). When analyzing these
models, we mostly rely on the assumption of ``slow-roll''
\cite{Stewart:1993bc, Sasaki:1995aw, Martin:2002vn} -- the slow
evolution of the inflaton field -- although fast-roll models have also
been considered \cite{Linde:2001ae}. Slow roll is an approximation
made to solve the perturbation equations, but not an absolute
requirement for an extended period of accelerated expansion. Even
though the scalar spectral index has arguably been measured by the
WMAP satellite\cite{Dunkley:2008ie} to be smaller than unity,
$n_s\approx0.96$, this constrains the slow-roll parameters
\begin{eqnarray}
  \epsilon\equiv-\frac{\partial_t{H}}{H^2}\,, \eta\equiv2\frac{\partial_\phi^2H}{H}\,, \zeta^2=\frac{1}{2}\frac{\partial_\phi H \partial_\phi^3H}{H^2}\,,
\end{eqnarray}
all in reduced Planck units $M_p^2=\frac{1}{8\pi G}=1$, only if they
are small. This was exactly the assumption made when solving the
perturbation equations, be it using Hankel functions
\cite{Stewart:1993bc}, the $\Delta N$ formalism \cite{Sasaki:1995aw,
  Starobinsky:1986fxa}, or the WKB approximation
\cite{Starobinsky:1982ee, Martin:2002vn}. If the slow-roll parameters
are large, then it is impossible to make any general statement about
their values from the measurement of $n_s$.

While it is true that $0<\epsilon<1$ for successful inflation, higher
order slow-roll parameters may be large, contrary to
intuition. Besides exploring unknown corners of inflationary model
space, there are many additional motivations to study models with
large $\eta$. For example, string theory motivated models of inflation
are typically characterized by large values of $\eta$
\cite{Kachru:2003sx}. Also, inflationary non-Gaussianities are
suppressed by powers of the slow-roll parameters
\cite{Maldacena:2002vr}, so that one might expect large
non-Gaussianities in fast-roll inflationary models.

As the problem of analyzing the behavior of fast-roll inflationary
models is technically very challenging, so far those models received
only limited attention. Essentially, there two ways to study this
largely unexplored realm. One of them is finding exact solutions to a
coupled system of non-linear differential equations. Another approach
is to solve these equations numerically. In our paper, we adopt both
approaches, providing an exact solution for a subset of initial
conditions as well as numerically charting the phase space of a new
class of inflationary models with large acceleration, i.e.  large
$\eta$. This allows us to study the behavior of both background and
perturbations in new and interesting regimes, where we find various
unexpected and even downright bizarre phenomena.

While our model can be made to agree with observations for a small
range of parameters, it possesses several most unusual features for
parameter ranges that are not compatible with observations. Some of
these features might seem counter-intuitive.  For example, according
to general lore, perturbation modes that exit the Hubble horizon,
freeze out and remain classical until the present epoch. We find that
is not true for a particular model of this family. The modes remain
quantum mechanical at all times and wavelengths, even long after they
cross the Hubble scale. Also, it is generally believed that the
interactions among the different modes of scalar perturbations are
suppressed by powers of the slow roll parameters, so that only
insignificant amounts of non-Gaussianity are created during
single-field inflation. Instead we show that it is the interaction
between different modes that defines the very form of the scalar power
spectrum. Other unusual features include the fact that tensor and
scalar modes can freeze out at different times and even unfreeze
during inflation. Finally, it turns out that for a specific member of
this family, the theory of scalar perturbations is described by a
quantum field theory in flat space time.

\begin{figure*}[t!]
  (a)\includegraphics[width=0.45\linewidth]{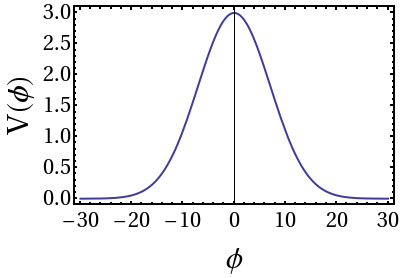}
  (b)\includegraphics[width=0.45\linewidth]{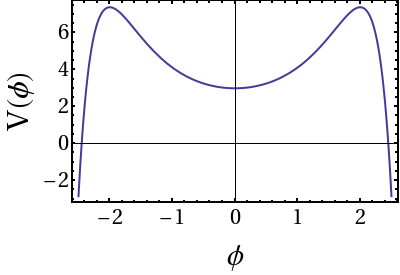}
  \caption{(a)$V(\phi)$ for $p=0.04$. Inflation takes place close to
    the top of the potential, rolling down from $\phi\approx0$ towards
    larger values of $|\phi|$. The potential is very shallow and turns
    negative only for $\phi>\frac{\sqrt{24}}{0.04}\approx122$,
    i.~e.~about $30$ e-folds after the end of inflation. For this
    model, the power spectrum of scalar perturbations has a spectral
    index that is compatible with observations,
    $n_s=1-p=0.96$. (b)$V(\phi)$ for $p=-2$. For negative $p$, the
    potential goes to $-\infty$ for $\phi\rightarrow\infty$. Inflation
    takes place rolling from small values of $|\phi|$ towards $\phi=0$
    and ends through tunneling. The scalar power spectrum possess some
    interesting features, see text.}
  \label{fig:potentialp2}
\end{figure*}

Our paper is organized as follows. In section
\ref{sec:most_curious_model} we present an explicit one-parameter
family of fast-roll models whose background equation is exactly
solvable for certain initial conditions. In section
\ref{sec:linear_perturbations} we discuss the spectrum of scalar and
tensor perturbations, which can be reliably computed despite the fact
that the slow-roll parameter $\eta$ may be larger than unity. In
section \ref{sec:p-2} we present an in-depth analysis of the
perturbations for a special member of the family of models, where the
scalar perturbations do not follow the usual behavior of oscillating,
being stretched to the horizon, and freeze-out, but instead obey an
exactly massless harmonic oscillator type equation. In section
\ref{sec:conclusions} we conclude and offer some ideas for future work
on this and similar classes of models.

\section{A most curious model}\label{sec:most_curious_model}
In this section we discuss in detail the shape of the potential, solve
the background dynamics and analyze the phase space for our class of
models.

\subsection{Shape of potential}\label{sec:shape_of_potential}

\begin{figure*}[th]
  (a)\includegraphics[width=0.45\textwidth]{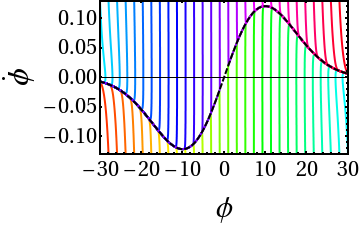}
  (b)\includegraphics[width=0.45\textwidth]{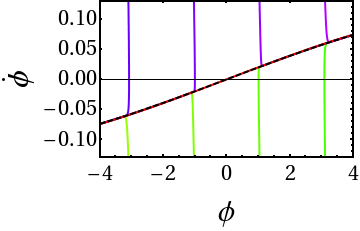}
  \caption{(a) Phase portrait for $p=0.04$. The black dashed line is
    the trajectory of the solution (\ref{eq:BackgroundSolution}). (c)
    Zoom of panel (a) with the hypothetical slow roll attractor in
    solid red. In this case, slow-roll is an attractor, owing to the
    fact that $\eta=-0.02$ is small.}
  \label{fig:phase_space0_04}
\end{figure*}

Consider a single real scalar field $\phi$ in a 4d
Friedman-Robertson-Walker universe
\begin{eqnarray}
  ds^2&=&dt^2-a^2(t)d\vec{x}^2\, ,
\end{eqnarray}
where the scale factor $a(t)$ is related to the Hubble parameter
$H(t)\equiv\frac{\dot{a}}{a}$. In units $M_p^2\equiv \frac{1}{8\pi
  G}=1$, the field's potential is given by
\begin{eqnarray}\label{eq:potential}
  V(\phi)&=&H_0^2 e^{-\frac{p}{4}\phi^2} \left(3-\frac{p^2}{8}\phi^2\right)\,,
\end{eqnarray}
where $H_0$ is some energy scale and $p$ is a real parameter. For
positive $p$, the minimum of the potential is located at
$\phi_{1,2}=\pm\frac{2\sqrt{6+p}}{p}$, with
$V(\phi_{1,2})=-H_0^2\frac{p}{2}e^{-\left(\frac{6}{p}+1\right)}$ and
its maximum is located at $\phi_3=0$, with $V(\phi_3)=3H_0^2$. For
negative $p$, the minima turn into maxima and the maximum turns into a
local minimum. (See Figure~\ref{fig:potentialp2}.) Note that for $p\le
-6$, the potential possess just a maximum and no minima, mimicking the
behavior of hill-top models \cite{Boubekeur:2005zm}. The potential is
negative for field values $|\phi|>\frac{\sqrt{24}}{|p|}$ independent
of the sign of $p$.
For positive $p$, we use the form of the potential
\begin{eqnarray}
  V&=&H^2(3-\epsilon)\,,
\end{eqnarray}
which is exactly equivalent to (\ref{eq:eom}), to see that the
potential turns negative when $\epsilon=3$, i.e. when
$\phi=\phi_-\equiv\frac{\sqrt{24}}{p}$.  This occurs $\Delta
N=\frac{1}{p}\ln 3$ e-foldings after the end of inflation. Thus, if
the potential (\ref{eq:eom}) holds for $\phi>\phi_-$, then the
universe would subsequently rapidly contract without entering an AdS
phase \cite{Felder:2002jk}.

If on the other hand $p<0$, inflation happens close to the local
minimum near the origin. The field $\phi$ slowly rolls down the hill
towards the minimum of its potential, so that inflation never ends in
the classical approximation. The model $p=-2$ was first discussed by
\cite{Easther:1995pc, Starobinsky:2005ab} who reach somewhat different
conclusions.

If quantum effects are taken into account, this conclusion
changes. For $p<0\neq-2$, Hawking-Moss tunneling\footnote{For $p=-2$,
  the Hawking-Moss rate is undefined since stochastic eternal
  inflation is absent, see Section~\ref{sec:p-2}.}
\cite{Hawking:1981fz} through the potential barrier will terminate the
inflationary phase in any given Hubble patch.  Still, the average
length of the inflationary stage will be enormously large --- the
inverse Hawking-Moss tunneling rate (per Hubble volume $H_0^{-3}$) is
exponentially suppressed
\begin{equation}
  t_{\rm inf}\sim{}H^{-1}_0\exp\left(\kappa\frac{M_P^2}{H_0^2}\right)\,,
\label{eq:HawkingMossTime}
\end{equation}
where $\kappa$ is a numerical factor of $O(1)$ and $H_0\ll M_p$.

Another issue is that the true minima of the potential are located at
$\phi\to\pm\infty$, $V=-\infty$. Because it is well-known
\cite{Felder:2002jk} that the universe collapses rapidly for negative
potentials, we need to avoid tunneling away from the local minimum to
negative values of the potential Thus we should define the potential
piece-wise as e.g.
\begin{eqnarray}\label{eq:potentialp-2}
  V&=&\left\{\begin{array}{ll}H_0^2e^{-\frac{p}{4}\phi^2}\left(3-\frac{p^2}{8}\phi^2\right),&|\phi|<\phi_*\\
   H_0^2e^{-\frac{p}{4}\phi_*^2}\left(3-\frac{p^2}{8}\phi_*^2\right),&|\phi|>\phi_*, \end{array} \right.
\end{eqnarray}
where $\phi_*$ is any arbitrary position between the zero crossings
\begin{eqnarray}
|\phi_*|&<\, \frac{\sqrt{24}}{|p|}.
\end{eqnarray}
This enables us to safely avoid tunneling to negative potential
energies. However we point out that even though such tunneling events
are rare, at late times the universe would be not necessarily be
sufficiently flat. Curing this would require a second phase of
inflation.

The universe would be automatically homogeneous if we introduced a
coupling of $\phi$ to a second field $\chi$ which is massive during
inflation and only becomes dynamically important after inflation,
i.e. once $\phi$ settled near its minimum $\phi=0$, similar to
waterfall models. 

\subsection{Classical background evolution}
The equations of motion for $\phi$ and the metric are of the usual form
\begin{eqnarray}\label{eq:eom}
  \ddot{\phi}+3H\dot{\phi}+\partial_\phi V&=&0\, ,\nonumber\\
  H^2=\frac{1}{3}\left(\frac{1}{2}\dot{\phi}^2+V\right)\, &,& \dot{H}=-\frac{1}{2}\dot{\phi}^2\, ,
\end{eqnarray}
where $\dot{\phi}\equiv\partial_t \phi$.  After switching the
independent variable from $t$ to the number of e-folds $N$ ($dN=Hdt$,
with $N=0$ at the beginning of inflation, $N>0$ at later times --
opposite to the standard definition of $N$), we find
\begin{eqnarray}
  \partial_N^2 \phi + (3 - \epsilon)\partial_N \phi + \frac{\partial_\phi V}{H^2}&=&0\,,\nonumber\\
  H^2\left(3-\frac{1}{2}(\partial_N\phi)^2\right)=V&,&\partial_N \ln H=-\frac{1}{2}(\partial_N\phi)^2\, .\quad
\end{eqnarray}
If we want the universe to undergo a period of inflation/accelerated
expansion, then we must have
\begin{eqnarray}\label{eq:inflation_condition}
  0 < \ddot{a}=a\left(\dot{H}+H^2\right)&=&aH^2\left(1-\epsilon\right)\,.\quad
\end{eqnarray}
Inflation thus happens whenever $0<\epsilon<1$.

\begin{figure*}[ht]
  (a)\includegraphics[width=0.45\textwidth]{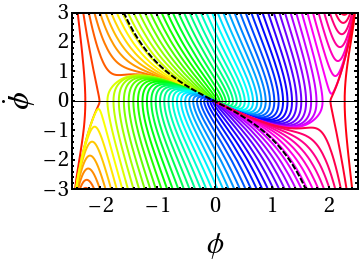}
  (b)\includegraphics[width=0.48\textwidth]{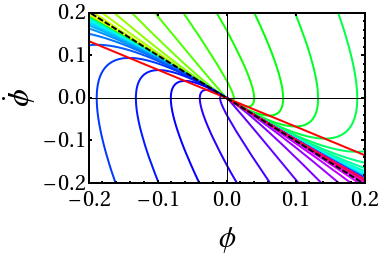}
  \caption{(a) Phase protrait for $p=-2$. The black dashed line is the
    trajectory of the solution (\ref{eq:BackgroundSolution}). (c) Zoom
    of panel (a) with the hypothetical slow roll attractor in solid
    red. Obviously, slow-roll is not an attractor in this model, in
    agreement with the fact that $\eta=1$ is large.}
  \label{fig:phase_space-2}
\end{figure*}

Of course, it is quite impossible to solve this set of differential
equations for general initial conditions. One exact solution for the
subset of initial conditions (see the appendix for details)
\begin{eqnarray}
  \phi(t=0)&=&\frac{2\sqrt{2\epsilon_0}}{p}\,,\nonumber\\
  \dot{\phi}(t=0)&=&H_0\sqrt{2\epsilon_0}e^{-\frac{\epsilon_0}{p}}\,,\nonumber\\
  H(t=0)&=&H_0 e^{-\frac{\epsilon_0}{p}}\,,
\end{eqnarray}
where we keep $a_0$ explicit for use in the next section, has the form
\begin{eqnarray}
  a=a_0 e^{N}\,, H=H_0e^{-\frac{\epsilon_0}{p}e^{pN}}\,, \epsilon=\epsilon_0e^{pN}\,, \phi=\frac{2\sqrt{2\epsilon_0}}{p}e^{\frac{pN}{2}}\, ,\,
\label{eq:BackgroundSolution}
\end{eqnarray}
Thus this solution has
\begin{eqnarray}
  \label{eq:SolutionProperties}
  H(\phi)= H_0 e^{-p\phi^2/8} \quad &\text{and}& \epsilon = \frac{p^2}{8}\phi^2 \,,
\end{eqnarray}
with initial values for $\phi, \dot{\phi}$, and $H$ fixed by the
choice of $\epsilon_0$ for a model with given $p$. Note that the
parameter $\eta$ is given by
\begin{eqnarray}\label{eq:eta_p}
  \eta&=&-\frac{p}{2}+\epsilon_0 e^{pN}=-\frac{p}{2}+\epsilon\, ,
\end{eqnarray}
that is, $|\eta|\approx \frac{p}{2}$ during most of the inflationary
period which in principle can be arbitrarily large.

There are $2$ qualitatively different regimes in this family of
models, $p$ positive and $p$ negative.

For $p>0$, $\epsilon$ grows with time, and inflation ends when
$\epsilon=1$. The field starts near the global maximum $\phi=0$, rolls
away and accelerates until inflation ends at
$\phi_{\text{end}}=\frac{2\sqrt{2}}{p}$. Defining $N_f$ to be the end
of inflation fixes $\epsilon_0\equiv e^{-pN_f}$.  If we arbitrarily
fix $N=0$ to be the onset of inflation, then $N_f$ is also the number
of inflationary e-folds.

The phase space, shown in Figure~\ref{fig:phase_space0_04} for
$p=0.04$, has a single unstable fixed point at the
origin\footnote{This value of $p\approx0.04$ leads to a power spectrum
  of scalar perturbations compatible with observations, see
  Section~\ref{sec:scalar_power}.}. The inflationary attractor is the
slow roll attractor (solid red line) and seems identical to the
background solution (\ref{eq:BackgroundSolution}) (dashed black
line). There are no fixed point attractors shown as the universe will
collapse as soon as the potential turns negative for
$|\phi|>\frac{\sqrt{24}}{p}$ (not shown).

If $p<0$, then $\epsilon$ decreases with time.  As we assume that
inflation starts at $N=0$, we take $\epsilon_0=1$ for negative $p$.
The time evolution according to (\ref{eq:BackgroundSolution}) does not
lead to an end of inflation at any particular final number of efolds
$N_f$. The field will reach the local minimum only in the infinite
future. The field may start rolling from
$|\phi|>|\phi_{\mathrm{crit}}|=|\frac{2\sqrt{2}}{p}|$, but only for
values $|\phi|<|\phi_{\mathrm{crit}}|$ will inflation actually
begin. Note that we can make $\eta$ large by making $p$ large and
negative and still have an infinite number of efolds of
inflation. However, for $p<-6$, the background solution
(\ref{eq:BackgroundSolution} )is unstable. It corresponds to $\phi$
starting its evolution with just enough momentum to roll up the hill
and come to a rest on the top.

The phase portrait, shown in Figure~\ref{fig:phase_space-2} for $p=-2$,
has a single fixed point attractor at the origin, corresponding to the
local minimum of the potential. Unlike chaotic $m^2\phi^2$ inflation,
there are no circular trajectories around the attractive fix-point. In
other words, the field does not oscillate around the local
minimum. The local maxima of the potential at $\phi=\pm2$ manifest
themselves as unstable points in the phase space diagram. For
$|\phi|>2$, the trajectories run to regions where the potential is
negative, leading to the collapse discussed by \cite{Felder:2002jk}
(not shown in the Figure).  At late times, the solution
(\ref{eq:BackgroundSolution}) obviously approaches the attractor
solution. Notice that the latter (dashed black line) differs from the
slow-roll attractor (solid red line)
\begin{eqnarray}\label{eq:attractors}
  \dot{\phi}_{\text{SR}}&=&-\frac{\partial_\phi V}{3H^2}\approx \frac{H_0}{12}p(6+p)\phi=-\frac{2}{3}H_0\phi\,,\nonumber\\
  \dot{\phi}_{\text{exact}}&=&\frac{p}{2}H_0\phi e^{-p\frac{\phi^2}{8}}\approx \frac{p}{2}H_0\phi=-H_0\phi\,,
\end{eqnarray}
demonstrating again that we are not dealing with a model of slow-roll
inflation, see Figure~\ref{fig:phase_space-2}(b). Thus, it is
justified to use (\ref{eq:BackgroundSolution}) as an asymptotic
solution when dealing with the late-time evolution and focus on its
properties.

\section{Linear perturbations}\label{sec:linear_perturbations}
In this section we describe the behavior of linear scalar and tensor
perturbations around the attractor solution found in the previous
section and compute their power spectra for generic values of $p$. As
it turns out, scalar and tensors freeze out at widely different times
for large values of $|p|$.

\subsection{Scalar power spectrum}\label{sec:scalar_power}
Using the formalism of \cite{Stewart:1993bc}, it is
straightforward to examine the scalar perturbations of the metric
\begin{eqnarray}
  u_k^\pprime(\tau) + \left(k^2 - \frac{z^\pprime}{z}\right)u_k(\tau)&=&0\, ,\label{eq:pert_eq}
\end{eqnarray}
where
\begin{eqnarray}
u_k&\equiv&-\frac{a\dot{\phi}}{H}\left(\Psi-\frac{H}{\dot{\phi}}\delta\phi\right).
\end{eqnarray}
Here, $\delta\phi$ is the perturbation of the inflaton field $\phi$,
$\Psi$ is the scalar perturbation of the spatial part of the metric,
$\tau$ is conformal time, $u_k^\prime\equiv\partial_\tau
u_k=a\partial_t u_k$ and $z\equiv \frac{a\dot{\phi}}{H}\equiv
a\sqrt{2\epsilon}$ for any single-field inflation model. We assume
that we are at late times and thus evolve along the attractor
solution, coinciding with the exact solution
(\ref{eq:BackgroundSolution}), to obtain
\begin{eqnarray}\label{eq:horizon_def}
  z&=&\sqrt{2\epsilon_0}a_0 e^{\left(\frac{p}{2}+1\right)N}\,\nonumber \\  {\rm and}\quad\quad& \nonumber \\
\frac{z^\pprime}{z}&=&a^2H^2\left(\frac{p}{2}+1\right)\left(\frac{p}{2}+2-\epsilon\right)\, ,\,\,
\end{eqnarray}
leading to the equations of motion for scalar perturbations $u_k$
\begin{eqnarray}\label{eq:scalar_mode}
  u_k^\pprime+\left[k^2 -a^2H^2\left(\frac{p}{2}+1\right)\left(\frac{p}{2}+2-\epsilon\right)\right]u_k&=&0\,.\,\,
\end{eqnarray}

\begin{figure*}[ht]
  (a)\includegraphics[width=0.45\textwidth]{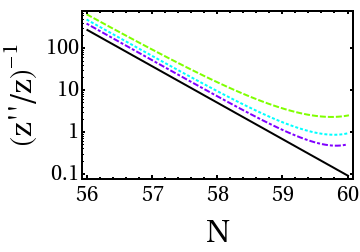}
  (b)\includegraphics[width=0.45\textwidth]{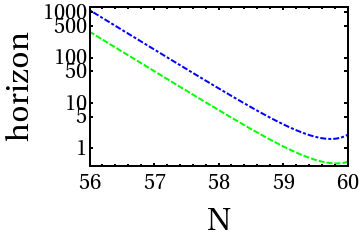}
  \caption{(a) Comoving horizon $(\frac{z^{\pprime}}{z})^{-1}$ for the
    scalar perturbations as a function of the number of e-folds $N$,
    see (\ref{eq:horizon_def}) for the exact definition. Notice that
    there is a maximum comoving wavenumber $k_*$ that will freeze out,
    depending on the value of $p$, and that some modes will leave the
    horizon and come back into the horizon {\it during} inflation,
    i.e. while $\epsilon<1$. $p=1$ (dashed green), $p=1.5$ (dotted
    blue), $p=2$ (dot-dashed purple), $p=-1$ (solid black).
    (b) Comparison of the comoving horizon for the scalar
    $(\frac{z^\pprime}{z})^{-1}$ (dashed green) and tensor
    $(\frac{a^\pprime}{a})^{-1}$ (dot-dashed blue) perturbations. For
    $p=2$, tensors freeze out about $1$ e-fold after the scalars. Note
    that inflation starts at $N=0$ and ends (for $p>0$) at $N=60$.}
  \label{fig:zpprime_max}
\end{figure*}

Before we attempt to solve this equations, we note that
$\frac{z^\pprime}{z}$ possesses a local maximum before the end of
inflation, implying that some modes might reenter the horizon during
inflation\footnote{If $k^2>\frac{z^\pprime}{z}$, then the modes behave
  like free oscillators, and if $k^2<\frac{z^\pprime}{z}$, the modes
  will freeze out.} (see Figure \ref{fig:zpprime_max}) albeit on
scales much smaller than observable in the CMB today. For
$p<-2-\sqrt{3}$ and $p>-2+\sqrt{3}$, their wave numbers are given by
values close to $k_*=\frac{z^\pprime}{z}\left(N_*\right)$, with
\begin{eqnarray}
  N_*&=&\frac{1}{p}\ln\left(\frac{3+p\pm\sqrt{p^2+4p+1}}{2\epsilon_0}\right)\,,
\end{eqnarray}
determined by the condition that $\frac{\partial}{\partial
  N}\frac{z^\pprime}{z}\Big|_{N=N_*}=0$. Similar effects were observed
by \cite{Leach:2000yw, Leach:2001zf}.

Let us return to solving the mode equation (\ref{eq:scalar_mode})
where $\epsilon\ll1$ around the time that the relevant scales for the
CMB leave the horizon ($N=0\dots10$ e-folds, i.e. $50\dots60$ e-folds
before the end of inflation).  Because $\epsilon\ll1$, $H$ is
approximately constant, $H\simeq H_0$.  In this approximation, the
solutions are given by (using the Bunch-Davis vacuum
$u_k=\frac{1}{\sqrt{2k}}e^{-ik\tau}$ as initial conditions)
\begin{eqnarray}
  u_k(\tau)&=&e^{i(\frac{p}{4}+1)\pi} \frac{\sqrt{\pi}}{2} \sqrt{-\tau} H_{\frac{3+p}{2}}^{(1)}(-k\tau) \, ,
\end{eqnarray}
which are approximated at late times ($\tau\rightarrow0$), i.e. after
freeze-out, as
\begin{eqnarray}
  u_k(\tau)= -i 2^{\frac{p+1}{2}}\sqrt{\frac{-\tau}{\pi}}\Gamma\left(\frac{p+3}{2}\right) e^{i(\frac{p}{4}+1)\pi} (-k\tau)^{-\frac{p+3}{2}}\,.\quad
    \label{eq:Asymp}
\end{eqnarray}
Thus we can compute the power spectrum of scalar perturbations (we
insert explicit factors of $M_p$)
\begin{eqnarray}
  \mc{P}_S&=&\frac{2^{p-1}}{\pi^3}\left[\Gamma\left(\frac{p+3}{2}\right)\right]^2\frac{H_0^2}{M_p^2}\frac{a_0^p}{\epsilon_0}\left(\frac{k}{H_0}\right)^{-p}\,\\
  &=&\frac{2^{p-1}}{\pi^3} \frac{3.81\times 10^{56p}}{\epsilon_0}\left[\Gamma\left(\frac{p+3}{2}\right)\right]^2\left(\frac{H_0}{M_p}\right)^{p+2}\left(\frac{k/a_0}{\mathrm{Mpc}^{-1}}\right)^{-p}\,,\nonumber
\end{eqnarray}
where $\epsilon_0=e^{-p\Delta N}$ for a duration of inflation of
$\Delta N$ efolds, and $a_0$ is the scale factor at the onset of
inflation. The scalar spectral index $n_s-1=-p$ is
consistent\footnote{Also, after inflation ends in this model, the
  universe will stay in a expanding phase for another $\Delta
  N=16\dots110$ efolds. }with measurements \cite{Dunkley:2008ie} by
the WMAP satellite $0.93<n_s<0.99$ ($95\%$CL) for $0.01<p<0.07$,
leading to values of $H_0\approx 10^{-4}\dots 10^{-5}M_p$ for a scalar
amplitude of $A_s=2.44\times 10^{-9}$. As we show below, the ratio of
amplitude of the tensor spectrum and the scalar spectrum can be tuned
to be compatible with observations if inflation lasts sufficiently
long.

\subsection{Tensor power spectrum and freeze out}
The equations of motion for the tensor perturbations $v_k$ are given
exactly by
\begin{eqnarray}
  \label{eq:eom_vk}
  v_k^\pprime+\left[k^2-a^2H^2(2-\epsilon)\right]v_k&=&0\, ,
\end{eqnarray}
where $v_k\equiv a\psi_{k,\lambda}$, with the tensor part of the metric
perturbation $h_{ij}=\int\!\frac{d^3k}{(2\pi)^{3/2}}\sum_\lambda
\psi_{k,\lambda}(\tau) e_{ij}(k,\lambda)e^{ikx}$ and transverse
polarization tensor $e_{ij}$. They can be solved in an analogous
manner to the scalar perturbations, again assuming $\epsilon$ is tiny
at the times of interest, to obtain an exactly scale free spectrum
\begin{eqnarray}
  \mc{P}_T&=&\frac{H_0^2}{4\pi^2} k^0\, ,
\end{eqnarray}
and tensor-scalar ratio $r\equiv\frac{P_T}{P_S}$
\begin{eqnarray}
  r&=&\frac{\epsilon_0}{3.81\times10^{56p}\left[\Gamma\left(\frac{p+3}{2}\right)\right]^2}\frac{\pi}{2^{p+1}}\left(\frac{H_0}{M_p}\right)^{-p}\left(\frac{k/a_0}{\mathrm{Mpc^{-1}}}\right)^p\,,\quad
\end{eqnarray}
which is $r<0.4$ for inflation lasting $120$ efolds and $p=0.01$ and
$r<10^{-4}$ for inflation lasting $60$ efolds and $p=0.07$. The former
is marginally excluded by observations, whereas the latter is well
compatible with observations.

From the equations of motion, it is immediately clear that tensors
freeze out at a different time than scalars (see
Figure~\ref{fig:zpprime_max}).  Due to the difference between
$\frac{z^\pprime}{z}$ and $\frac{a^\pprime}{a}$, tensor and scalar
modes of the same comoving wavelength cross their respective horizons
at different times. Conversely, at each point in time, scalar and
tensor modes of wavenumber difference $\Delta k^2$ cross their
horizons, with
\begin{eqnarray}
  \sqrt{\Delta k^2}&=&\sqrt{\left|\frac{z^\pprime}{z}-\frac{a^\pprime}{a}\right|}=aH\sqrt{\frac{p}{2}\left(\frac{p}{2}+3-\epsilon\right)}\, .
\end{eqnarray}
Increasing $|p|$ leads to larger and larger differences in the
freeze-out time\footnote{Note that large $p$ are inconsistent with
  observations.}.

\section{Special case without freeze-out}\label{sec:p-2}

There are several models with specific values of $p$ that deserve
special attention as they show unexpected behavior of the
perturbation modes.  First of all, when $-2>p>-4$, the sign of
$\frac{z^\pprime}{z}$ for the background solution/attractor
flips. Thus the scalar perturbations obtain a positive mass squared
and consequently do not freeze-out.

\begin{figure*}
  (a)\includegraphics[width=0.45\textwidth]{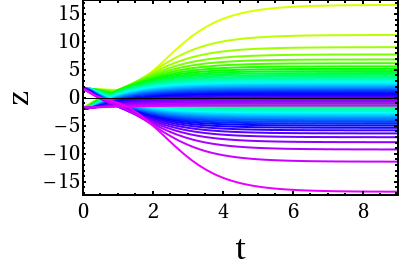}
  (b)\includegraphics[width=0.45\textwidth]{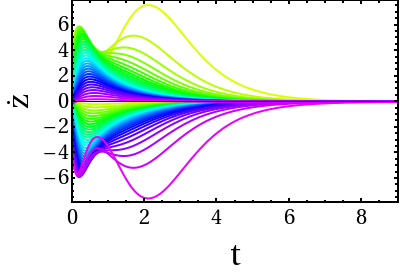}
  \caption{(a) $z=\sqrt{2\epsilon}a$ and (b) $\dot{z}$ as a function
    of the number of efolds $N$. Notice that independent of the
    initial values of $\phi, \dot{\phi}$, $z\approx\mathrm{constant}$
    after about 5 efolds, while inflation lasts exponentially many
    efolds. The color coding is the same as
    Figure~\ref{fig:phase_space-2}.}
  \label{fig:z_zprime}
\end{figure*}

Secondly, when $p=-2$, then $z=\sqrt{2\epsilon}a$ is identically
constant at late times (corresponding to small $\phi$) as we see from
the numerical analysis of the system for different initial conditions,
see Figure~\ref{fig:z_zprime}. As inflation lasts an exponentially
large number of efolds (owing to the exponentially suppressed rate of
tunneling out of the false minimum at $\phi=0$), we can use this
observation to study the generation of fluctuations assuming
$z=\mathrm{constant}$ along all phase space trajectories.
Thus, the equation of motion for $u_k$ simplifies to
\begin{eqnarray}
  u_k^\pprime+k^2 u_k&=&0\, ,
\label{eq:HarmonicOscillModes}
\end{eqnarray}
indicating that a) particle creation by the gravitational field is
absent and b) modes {\it never freeze out}, not even when the physical
wavelength of a given mode exceeds the Hubble scale.\footnote{Note
  that for the case $p=-2$ the asymptotic form (\ref{eq:Asymp}) does
  not hold.} 

We stress that by freeze-out we refer to the regime of the
perturbation equations (\ref{eq:pert_eq}) where the $k^2$ term becomes
sub-dominant compared to effective negative mass-square term. Although
there is only a finite amount of conformal time after a given mode
leaves the horizon, this does not make the mode classical even at
$\tau->0$ in our case, but only slows down the rate of oscillations in
physical time.

The absence of a quasi-classical regime implies that there is no
stochastic eternal inflation, as the modes $u_k$ never become
classical. This is easy to understand\footnote{We thank Latham Boyle
  for pointing this out.} when looking at the action
\cite{Maldacena:2002vr} for the curvature perturbations $\zeta$
\begin{eqnarray}
  S&=&\int d^4x \sqrt{-g} \epsilon g^{\mu\nu}\partial_\mu\zeta\partial_\nu\zeta\,.
\end{eqnarray}
For $p=-2$, the scale-factor dependence of $\sqrt{-g}$, $\epsilon$ and
$g^{\mu\nu}$ precisely cancel, making $\zeta$ ``live'' in Minkowski
space.

In ``ordinary'' inflationary models, the resulting path integral would
become classical as the action is growing exponentially due to the
scale-factor dependence. In this particular model, the action will not
grow like a power of $a$, thus the path integral
\begin{eqnarray}
  Z=\exp\left({\frac{iS}{\hbar}}\right)
\end{eqnarray}
will remain dominated by quantum effects.


Let us first understand why $z$ approaches constant value at late
times.  As explained above, for almost arbitrary initial conditions
$\phi(t=0)=\phi_0$, $\dot{\phi}(t=0)=\dot{\phi}_0$ the system reaches
the attractor trajectory in finite time. For $p=-2$ and at small $\phi
< M_P$ the attractor trajectory is given by the expression
\begin{equation}
  \dot{\phi}=-H_0\phi{}
  \label{eq:Attrp2}
\end{equation}

Solving the background FRW equations for the trajectory
(\ref{eq:Attrp2}), one can easily check that
\begin{equation}
  a\phi=\mathrm {constant}\,,
  \label{eq:Udef}
\end{equation}
in this regime. In other words, $\phi\sim{}a^{-1}$ at late times:
although $\phi$ becomes exponentially small, classically it takes an
{\it infinite} amount of time for the system to reach the false
minimum $\phi=0$.

 Also note that (\ref{eq:Udef}) can be derived without using the
 asymptotic expression for the attractor (\ref{eq:Attrp2}). Indeed,
 the equation of motion for the variable $U=a\phi$ is given by
\begin{equation}
  U''-\frac{a''}{a}U+a^4\frac{\partial V(U/a)}{\partial{}U}=0,
  \label{eq:1anottaken}
\end{equation}
where $^\prime$ denotes derivative w.r.t. conformal time $\tau$. Since
inflation lasts infinitely long, one can use the expansion in powers
of $1/a$ in order to find its asymptotic solution at late times.

To leading order in $1/a$, equation (\ref{eq:1anottaken}) acquires the
form
\begin{equation}
  U''=0\,.
\label{eq:U}
\end{equation}
Note that this equation is independent of $a$ and, as we will see
below, the equation for the Fourier modes $u_k$ also has non-trivial
$a$-independent form in the limit $1/a\to\infty$. This is what makes
our model special. Indeed, for a generic inflationary model one would
have
\begin{eqnarray}
  u_k''+k^2u_k+\left(m^2a^2-\frac{a''}{a}\right)u_k&=&0\,,
\end{eqnarray}
which can be approximated to
\begin{eqnarray}
  u_k''+k^2u_k+(m^2-2H^2)a^2 u_k&=&0\,,
  \label{eq:1ageneric}
\end{eqnarray}
to leading order in $1/a$. There, the $1/a$ expansion would only
illustrate the familiar fact that modes with effective mass $m^2>H^2$
get amplified by inflation, while modes with $m^2<H^2$ get
exponentially damped. In our model with $p=-2$, the effective mass
term proportional to $a^2$ cancels exactly.


Let us now discuss the dynamics of fluctuations around the background
$U=\mathrm{constant}$.  Since $\frac{z''}{z}=0$ in our case, the
leading effect influencing the dynamics of the modes $u_{k}$ is the
interaction between them.

To estimate the leading (third order in small perturbations) effect of
this interaction, it is necessary to work in the constant curvature
gauge fixed to second order in small perturbations \cite{Maldacena:2002vr}
\begin{eqnarray}
  &a\phi=U+u(\tau,x)\,,u=a\delta\phi(\tau,x)\,,h_{ij}=a^2\hat{h}_{ij},&\\
  &{\rm det}\hat{h}=1\,,\hat{h}_{ij}=\delta_{ij}+\tilde{\gamma}_{ij}+
  \frac{1}{2}\tilde{\gamma}_{il}\tilde{\gamma}_{lj}+\ldots\,,&\\
  &\tilde{\gamma}_{ii}=0\,,\partial_i{}\tilde{\gamma}_{ij}=0\,.&
  \label{eq:ConstCurvGauge}
\end{eqnarray}
Keeping only the leading order in $1/a$, one finds that the scalar
part of the ADM action for fluctuations has the form
\begin{equation}\label{eq:S_a_independent}
S=\int_{-1/H}^{0}{}d\tau{}\int{}d^3x\left(\frac{1}{2}(u')^2-\frac{1}{2}(\partial_x{}u)^2-W(u)\right),
\label{eq:Sadm}
\end{equation}
where
\begin{equation}
W(u)=h_u{}u^3,\ h_u=\frac{H_0^2}{4M_P^2}U\,,
\end{equation}
in the third order w.r.t. small fluctuations. Note that the action is
independent of the scale factor $a$ in the limit $a\rightarrow
\infty$, e.g. $|\tau|\ll H^{-1}$. The interaction terms do not get
red-shifted, neither do the occupation numbers for the modes $u_k$ nor
the effective temperature of the perturbations $T$. In other words,
since $U$ is expected to be constant at late times, we see that
\emph{in the limit $1/a\to{}0$, the physics of fluctuations on around
  this quasi-de Sitter space is described by the quantum field theory
  (\ref{eq:Sadm}) living in flat spacetime}.

Then, the question of initial conditions for the model becomes very
important for the model (\ref{eq:S_a_independent}). While it is
natural to assume $U=\mathrm{constant}$ as initial condition for the
background (trajectories corresponding to other choice of initial
conditions approach the attractor $U=\mathrm{constant}$ in finite
time), the initial conditions for the perturbation modes $u_{k}$ can
be chosen arbitrarily. Choosing a multi particle initial state such
that
\begin{equation}
  n_{k}=\left\langle \frac{1}{2k}\left(|u_{k}'|^{2}+k^{2}u_{k}^{2}\right)\right\rangle -\frac{1}{2}>1\,,
\end{equation}
and $\langle|u_{k}|^{2}\rangle=\frac{n_{k}}{2\omega_{k}}$ is large for
certain intervals of $k$, one can expect that the distribution $n_{k}$
will change with time, gradually approaching a thermal distribution
\begin{equation}
  n_{k}\sim\exp\left(-\frac{k^2}{T_{{\rm therm}}}\right).
\end{equation}
The final value of the thermalization temperature $T_{{\rm therm}}$
can be easily estimated: since the equations of motion following from
(\ref{eq:Sadm}) have an integral of motion and preserve the total
``energy'' of the system
\begin{equation}\label{eq:energy_pumped}
  E=\int d^{3}x\left(\frac{1}{2}(u')^{2}+\frac{1}{2}(\nabla u)^{2}+W(u)\right)+\frac{1}{2}(U')^2,
\label{eq:energy}
\end{equation}
the thermalization temperature is determined by the amount of
``energy'' (\ref{eq:energy_pumped}) in the system at $\tau\approx
H^{-1}$
\begin{equation}
T_{{\rm therm}}\sim E^{1/4},
\end{equation}
where $V\sim{}H_0^{-3}$ is the 3-volume of the initial Hubble
patch. Note that the energy (\ref{eq:energy}) is unbounded from below,
if only second and third order terms w.r.t. small fluctuations are
kept in the action (\ref{eq:Sadm}). This issue is resolved by taking
4th order terms into account. No higher order terms survive in the
$1/a\to{}0$ approximation.

The question of thermalization time scale is actually non-trivial. For
example, if we choose an ``initial'' state at $\tau\approx H^{-1}$
such that $n_k$ is peaked at some $k=k_i$ and zero at other $k$, a
careful analysis \cite{Micha:2004bv} shows that complete
thermalization takes at least
\begin{equation}
\tau_{{\rm therm}}\sim \left(\frac{T_{\rm therm}}{k_i}\right)^{2/5}.
  \label{eq:ThermalizationTimeProperEstimate}
\end{equation}

The physical reason for the long thermalization time scale is the
following.  The evolution of $n_k$ is described by a system of coupled
kinetic equations. Typically, when initial occupation numbers $n_k$
are large, the general solution this system gets quickly attracted to
(and spends a long time in) a self--similar regime of weak wave
turbulence \cite{Micha:2004bv}, characterized by a power law spectrum
\begin{equation}
  n_{k}\sim k^{-3/2}
\label{eq:WeakTurbSpectrum}
\end{equation}
and by a scaling behavior of occupation numbers with time.

Note on the other hand that the spectrum of perturbations has the
Rayleigh-Jeans form \cite{Podolsky:2005bw}
\begin{equation}
n_k\sim\frac{T}{\omega_k}\sim{}k^{-1}
\label{eq:RJSpectrum}
\end{equation}
in the infrared part, if ultimate thermalization is achieved.

The spectra (\ref{eq:WeakTurbSpectrum}) and (\ref{eq:RJSpectrum}) are
trivially related to the power spectrum of primordial perturbations:
\begin{equation}
k^{3}{\cal P}_{k}=k^{3}\left|\frac{u_{k}}{z}\right|^{2}=\frac{k^{3}n_{k}}{z^{2}\omega_{k}}.
\label{eq:PowSpecnk}
\end{equation}
However, the expression above holds interest for us if and only
if inflation comes to an end --- only in this case Eq. (\ref{eq:PowSpecnk})
is related to CMB anisotropies that we observe in the sky.

As discussed in Section~\ref{sec:shape_of_potential}, inflation can
only end due to tunneling from near the false minimum at $\phi=0$ to
the true minima at $\phi\to\pm\infty$. The duration of inflation can
be estimated as
\begin{eqnarray}
  \tau_{{\rm inf}}\sim\frac{1}{H_{0}a_{0}}\,.
\end{eqnarray}
The effective action for the fluctuations (\ref{eq:Sadm}) is only
valid during a finite amount\footnote{We thank the referee for
  reminding us of this fact.} of conformal time $-H^{-1}<\tau<0$. If
the thermalization time scale is longer than this period, the
perturbation spectrum will be given by
(\ref{eq:WeakTurbSpectrum}). The crucial point is that -- whether
there is enough conformal time for the system to reach full
thermalization or not -- the modes keep interacting even after they
leave the horizon, making the power spectrum change with time even for
superhorizon modes.

Let us also briefly discuss the issue of backreaction of
fluctuations $u_k$ on the background $U$. The presence of the third
order term in the ADM action (\ref{eq:Sadm}) leads to an instability
through growth of small fluctuations $u(\tau,x)$. Since the trilinear
vertex is of the form
\begin{equation}
  W_3\sim\frac{H_0^2}{M_P^2}Uu^3,
  \label{eq:trilinvertex}
\end{equation}
growth of $u$ leads to decrease of the background field
$U$. Ultimately, $u$ stops growing either when $U$ reaches $0$, so
that the trilinear vertex (\ref{eq:trilinvertex}) effectively
disappears, or when the quartic term $W_4\sim{}\frac{H_0^2}{M_P^2}u^4$
starts to dominate over the trilinear term in the effective potential
$W(u)$. While it would take an infinite amount of time for the
classical, unperturbed system to reach the minimum of the potential,
the background field will vanish in finite time if perturbations are
taken into account.

Finally, we would like to remark that the ADM action (\ref{eq:Sadm})
above contains only terms corresponding to interaction between scalar
modes. Strictly speaking, interaction between tensor and scalar modes
should also be taken into account, since it does not vanish in the
limit $a\to\infty$: namely, the term corresponding to decay of one
scalar into two tensors
\begin{equation}
W_{3,{\rm T}}\sim{}H_0^2{}Uuv^2
\end{equation}
is of the same order of magnitude as $W_3$ in the limit
$a\to\infty$. This fact has several important consequences. First of
all, since tensors do freeze out even for $p=-2$, super-horizon tensor
modes generate an effective term in the ADM effective action which is
linear in $u$ and $U$, leading to a faster decay of the background
field $U$.On the other hand, scalar fluctuations $u$ also effectively
decay into tensor modes with $k>aH$. We leave the study of this issue
for future work.

\section{Conclusion and Outlook}\label{sec:conclusions}

In this paper we discuss the behavior of the family of inflationary
models
\begin{eqnarray}
\label{eq:Vconclusion}
  V&=&H_0^2M_P^2 e^{-\frac{p}{4}\frac{\phi^2}{M_P^2}}\left(3-\frac{p^2}{8}\frac{\phi^2}{M_P^2}\right)\,,
\end{eqnarray}
with arbitrary $p$ and $H_0<M_{\mathrm{pl}}$.

Exploring the phase space of this family of models for large $|p|$
reveals the existence of a non-slow-roll attractor which at late times
can be approximated by
\begin{eqnarray}
  \dot{\phi}&=&\frac{p}{2}H_0\phi\,.
\label{eq:nonslowrollattrlate}
\end{eqnarray}
For small values of $|p|$ this attractor coincides with the slow-roll
attractor, see equation (\ref{eq:attractors}).

For all values of $|p|$, this family of models is observationally
ruled out. However, they turn out to exhibit many fascinating
features, some quite counterintuitive. The reason why the standard
lore based on the slow roll approximation fails in this case is that
the value of $p$ essentially determines the value of the slow roll
parameter $\eta=2\frac{\partial_\phi^2 H}{H}$, so that larger values
of $|p|$ correspond to larger values of $\eta$, see equation
(\ref{eq:eta_p}) .

Table \ref{tab:p_overview} show the physical properties of our family
of models for different values of the parameter $p$.
\begin{table}
  \begin{tabular}{|c|c|}
    \hline
    {\bf value of} $\mathbf{p}$ & {\bf model features}\\
    \hline
    $-\infty < p < -6$ & widely separated freeze-out\\
    & $\infty$ long inflation for one initial condition\\
    & $|\eta|>1$ \\
    \hline
    $-6<p<-4$ & widely separated freeze-out\\
    & $|\eta|>1$\\
    & $\infty$ long classical inflation\\
    \hline
    $-4<p<-2$ & no freeze out\\
    & $|\eta|>1$\\
    & $\infty$ long classical inflation\\
    \hline
    $p=-2$ &  no freeze out\\
    & no gravitational particle production\\
    & no stochastic eternal inflation\\
    & non-suppressed mode interactions\\
    & (pre)thermalization\\
    & dual to a QFT in flat space\\
    & $|\eta|\approx 1$\\
    & $\infty$ long classical inflation\\
    \hline
    $-2<p<-2+\sqrt{3}$ & modes reenter the horizon\\
    &$\infty$ long classical inflation\\
    \hline
    $-2+\sqrt{3}<p<0$& $\infty$ long classical inflation\\
    \hline
    $p=0$ & exact de Sitter space-time\\
    \hline
    $0<p<0.01$ & finite duration of inflation\\
    \hline
    $0.01<p<0.07$ & compatible with observations\\
    \hline
    $0.07<p<\infty$ & finite duration of inflation\\
    \hline
  \end{tabular}
  \caption{Behaviour of our model $V=H_0^2
    e^{-\frac{p}{4}\phi^2}(3-\frac{p^2}{8}\phi^2)$ for different
    values of $p$.}
  \label{tab:p_overview}
\end{table}
The most interesting physics takes place in the model with $p=-2$ (see
the Fig. 2b). First of all, to leading order in a $1/a$ expansion, the
scalar perturbations in this model turn out to be described by a
quantum field theory in flat spacetime (\ref{eq:Sadm}) . The linear
scalar perturbations are free massless harmonic oscillators and thus
never freeze out\footnote{In fact, models with $-4<p<-2$ are also
  characterized by the absence of freeze out. Here, the scalar
  perturbations are harmonic oscillators with time-varying but
  strictly positive masses. We leave a thorough examination of this
  issue to future work.}. Gravitational particle production,
stochastic kicks uplifting the background value of the inflaton field
and a regime of stochastic eternal inflation are completely absent.

If the initial state for the scalar fluctuations $u_k=a\delta\phi_k$
has a non-zero occupation number for a range of momenta, one finds
that the corresponding occupation numbers $n_k$ typically approach a
spectrum of form (\ref{eq:WeakTurbSpectrum}) due to interactions
between the modes $u_k$, with a spectral index
$n_s=\frac{3}{2}$. Interestingly, the effects of the interaction
between the different modes $u_k$ do not get redshifted away.

As a curious aside, we notice that the shape of the potential
(\ref{eq:potential}) is strongly reminiscent of the SUGRA form
\begin{eqnarray}
  V_{\mathrm{SUGRA}}&=&e^K\left(K^{ij} D_iW D_{\bar{j}}\bar{W} - 3|W|^2\right)\, ,
\end{eqnarray}
where the K\"ahler potential $K$ is a real function of the complex
fields $\phi^i, \phi^{\bar{i}}$, the superpotential $W$ is a
holomorphic function of the complex fields $\phi^i$, the covariant
derivative $D_iW=\partial_{\phi^i} W + \partial_{\phi^i}K W$ and
$K^{i\bar{j}}=(\partial_{\phi^{i}\phi^{\bar{j}}} K)^{-1}$. Assuming
the K\"ahler potential $K=-\frac{q}{4}\phi\bar{\phi}$ and the
superpotential $W=H_0=\mathrm{constant}$, the potential can be
computed to
\begin{eqnarray}
  V_{\mathrm{SUGRA}}&=&-H_0^2 e^{-\frac{q}{4}|\phi|^2}\left(3+\frac{q}{4}|\phi|^2\right)\, .
\end{eqnarray}
For $q=-2$, this agrees with (\ref{eq:Vconclusion}) for $p=-2$ up to
an overall sign
\begin{eqnarray}
  V_{\mathrm{SUGRA}}^{q=-2}&=&-V^{p=-2}\, ,
\end{eqnarray}
possibly indicating that this model may be embedded into string
theory.

Finally, we remark that it might be possible to have both modes
without freeze-out and a scalar power spectrum with $n_s\approx{}0.96$
for a wider class of models, where the attractor solution has
$z=\mathrm{constant}$ instead of $z^\prime=\mathrm{constant}$. In this
case, it is immediately clear that the equations of motion for the
perturbations $u_k$ again simplify to those of a free harmonic
oscillator. What is not clear is whether in these models inflation
actually takes place: the inflationary condition from
(\ref{eq:inflation_condition}) corresponds to a highly non-trivial
differential equation for the Hubble parameter $H$. We leave a
thorough examination of this class of models for future work.

\section*{Acknowledgements}
We are grateful to Neil Barnaby, Dick Bond, Latham Boyle, Zhiqi Huang,
Dominik Schwarz and Alexei Starobinsky for useful discussions. GDS, DIP
and PMV have been supported by grants from the US DOE and NASA to the
particle-astrophysics theory group at CWRU.

\begin{appendix}

\section{ADM action for scalar fluctuations in the case $p=-2$}

In this Appendix we explicitly derive the 3rd order ADM action
(\ref{eq:Sadm}) for the fluctuations $u=a\delta\phi$ in the case
$p=-2$ (for the sake of convenience, we will denote the fluctuations
$\delta\phi$ here by $\vphi$).

Calculated in the constant curvature gauge (\ref{eq:ConstCurvGauge}),
the 3rd order ADM action has the form \cite{Maldacena:2002vr}
\begin{eqnarray}
S_3&=&\int{}d\tau{}d^3 x a^4\left(-\frac{\dot{\phi}}{H}\vphi\dot{\vphi}^2-a^{-2}\frac{\dot{\phi}}{4H}\vphi
(\partial\vphi)^2-\right.\\ \nonumber
& &-\dot{\phi}\partial_i\chi\partial_i\vphi+\frac{3\dot{\phi}^3}{8H}\vphi^3-\frac{\dot{\phi}^5}{16H^3}\vphi^3-\frac{\dot{\phi}V''}{4H}\vphi^3-
\\ \nonumber
& &-\frac{V'''}{6}\vphi^3+\frac{\dot{\phi}^3}{4H^2}\vphi^2\dot{\vphi}+\frac{\dot{\phi}^2}{4H}\vphi^2\partial^2\chi+\\ \nonumber
& &\left. +\frac{\dot{\phi}}{4H}(-\vphi\partial_i\partial_j\chi\partial_i\partial_j\chi+\vphi\partial^2\chi\partial^2\chi)\right)
\label{eq:ADM3canonical}
\end{eqnarray}
where $\chi$ satisfies the equation
\begin{equation}
\partial^2\chi=\frac{\dot{\phi}^2}{2H^2}\frac{d}{dt}\left(-\frac{H}{\dot{\phi}}\vphi\right).
\end{equation}

To estimate how different terms in (\ref{eq:ADM3canonical}) behave at
$a\to\infty$, we note that in this regime all trajectories in the
phase space approach the non-slow roll attractor
$\dot{\phi}=-H_0\phi$. By definition, the variable $u$
\begin{equation}
\vphi=\frac{u}{a}\sim\frac{1}{a},
\end{equation}
since $u$ remains finite on the attractor. Therefore,
\begin{equation}
\dot{\delta\phi}=\frac{u'}{a^2}-\frac{Hu}{a}\approx{}-\frac{Hu}{a}
\end{equation}
to leading order in $1/a$ (here prime denotes the usual derivative
with respect to conformal time $\tau$). Also note that in the vicinity
of the attractor $\dot{\phi}=-H_0\phi$ the scalar $\chi$ evolves
according to
\begin{equation}
\partial^2\chi=\pm\frac{Uu'}{2a^3},
\end{equation}
so that $\chi$ has at most the order $1/a^3$ as $a\to\infty$.

Analyzing the 3rd order ADM action (\ref{eq:ADM3canonical}) term by
term, we find that\footnote{For briefness, we only focus on the case
  $\phi<0$, $\dot{\phi}>0$, generalization to the case $\phi>0$,
  $\dot{\phi}<0$ is straightforward.}
\begin{equation}
-a^4\frac{\dot{\phi}}{H}\vphi\dot{\vphi}^2=-\frac{H^2_0}{4}Uu^3\sim{}O(1),
\label{eq:1term}
\end{equation}
\begin{equation}
-a^4a^{-2}\frac{\dot{\phi}}{4H}\vphi(\partial\vphi)^2\sim{}O(a^{-2}),
\end{equation}
\begin{equation}
-a^4\dot{\phi}\partial_i\chi\partial_i\vphi\sim{}O(a^{-1}),
\end{equation}
\begin{equation}
a^4\frac{3\dot{\phi}^3}{8H}\vphi^3\sim{}O(a^{-2}),
\end{equation}
\begin{equation}
-a^4\frac{\dot{\phi}^5}{16H^3}\vphi^3\sim{}O(a^{-4}),
\end{equation}
\begin{equation}
-a^4\frac{\dot{\phi}V''}{4H}\vphi^3\sim{}O(1),
\label{eq:2term}
\end{equation}
\begin{equation}
-a^4\frac{V'''}{6}\vphi^3\sim{}O(1),
\label{eq:3term}
\end{equation}
\begin{equation}
a^4\frac{\dot{\phi}^3}{4H^2}\vphi^2\dot{\vphi}\sim{}O(a^{-2}),
\end{equation}
\begin{equation}
a^4\frac{\dot{\phi}^2}{4H}\vphi^2\partial^2\chi\sim{}O(a^{-3})
\end{equation}
and finally
\begin{equation}
a^4\frac{\dot{\phi}}{4H}(-\vphi\partial_i\partial_j\chi\partial_i\partial_j\chi+\vphi\partial^2\chi\partial^2\chi)\sim{}O(a^{-3}).
\end{equation}
As we find, only three terms (\ref{eq:1term}), (\ref{eq:2term}) and
(\ref{eq:3term}) survive out of 10 terms present in
(\ref{eq:ADM3canonical}) when $a\to\infty$. Moreover, one can
explicitly show that the terms (\ref{eq:2term}) and (\ref{eq:3term})
cancel each other, so that the term ((\ref{eq:1term})) is the only one
that contributes to the effective potential $W(u)$ in
(\ref{eq:Sadm}).

\section{Exact background solution and perturbations for a subset of initial conditions}
By definition, the ``slow-roll'' parameter $\epsilon$ is given by
\begin{eqnarray}
  \epsilon&=&-\frac{d\ln H}{dN}\, ,
\end{eqnarray}
where $N$ is the number of e-folds from the end of inflation,
$dN=Hdt$, where $t$ is physical time.

Useful formula are
\begin{eqnarray}
  a&=&e^{N}\, ,\\
  \frac{\partial\phi}{\partial N}&=&\sqrt{2\epsilon}\, ,\\
  V&=&H^2(3-\epsilon)\, ,\\
  \eta&=&-\frac{1}{2\epsilon}\frac{\partial\epsilon}{\partial N}+ \epsilon\, ,
\end{eqnarray}
which are exact equations with no assumptions about slow-roll. Let us
examine the one-parametric family of models
$\epsilon=\epsilon_0e^{pN}$.

\subsection{Background}
It is straightforward to integrate the above equations to obtain
\begin{eqnarray}
  H&=&H_0e^{-\frac{\epsilon_0}{p}e^{pN}}\, ,\\
  \phi&=&\frac{2\sqrt{2\epsilon_0}}{p}e^{\frac{pN}{2}}\, ,\\
  \Rightarrow H&=&H_0 e^{\frac{-p}{8}\phi^2}\, ,\\
  \epsilon&=&\frac{p^2}{8}\phi^2\, , \\
  \eta&=&-\frac{p}{2}+\epsilon_0e^{pN}\, ,\\
  V(\phi)&=&H_0^2 e^{\frac{p}{4}\phi^2}\left(3-\frac{p^2}{8}\phi^2\right)\, ,
\end{eqnarray}
where $H_0$ is a free integration constant. $|\eta|$ is arbitrarily
large for large $|p|$, yet inflation proceeds as
$0<\epsilon=\epsilon_0e^{|p|N}<1$. However, the potential has extrema
at $\frac{\partial V}{\partial \phi}=0$
\begin{eqnarray}
  \phi_1=0&,&V(\phi_1)=3H_0^2\,,\\
  \phi_{2/3}=\pm\frac{2\sqrt{6-p}}{p}& ,&V(\phi_{2/3})=H_0^2\frac{p}{2}e^{\frac{6}{p}-1}\, .
\end{eqnarray}

\subsection{Perturbations}
The perturbation equations for scalars/ tensors are given by
\cite{Stewart:1993bc}
\begin{eqnarray}
  u_k^\pprime+\left(k^2-\frac{z_{S,T}^\pprime}{z_{S,T}}\right)u_k&=&0\, ,
\end{eqnarray}
where $\prime$ denotes derivative with respect to conformal time
$d\tau=\frac{dt}{a}$, and $z_S=a\sqrt{2\epsilon}$ for scalar and
$z_T=a$ for tensor perturbations, where $a$ is the scale factor of the
FRW metric. Converting derivatives w.r.t. to $\tau$ to derivatives
w.r.t. $N$, we use
\begin{eqnarray}
  \partial_{\tau}&=&a\partial_t=aH\partial_N\, ,\\
  \partial_{\tau}^2&=&a^2H^2\partial_N^2+aH (\partial_N aH)\partial_N\,\nonumber\\
  &=&a^2H^2\left(\partial_N^2+(1-\epsilon)\partial_N\right)\, ,
\end{eqnarray}
to rewrite $\frac{z_{S,T}^\pprime}{z_{S,T}}$
\begin{eqnarray}
  \frac{a^\pprime}{a}&=&\frac{1}{a}\left(a^2H^2\partial_N^2 a + a^2H^2(\epsilon-1)\partial_N a\right)\, ,\\
  \frac{a''}{a}&=&a^2H^2(2-\epsilon)\,,\\
  z&=&\sqrt{2\epsilon}a=a_0\sqrt{2\epsilon_0}e^{-\left(\frac{p}{2}+1\right)N}\,,\\
  \frac{z^\pprime}{z}&=&a^2H^2\left(\frac{p}{2}+1\right)\left(\frac{p}{2}+2-\epsilon\right)\,,
\end{eqnarray}
and to obtain for the tensor modes $v_k$ and scalar modes $u_k$
\begin{eqnarray}
  u_k^\pprime + \left[k^2-a^2H^2\left(\frac{p}{2}+1\right)\left(\frac{p}{2}+2-\epsilon\right)\right] u_k&=&0\,,\quad\,\,\\
  v_k^\pprime + \left(k^2-a^2H^2(2-\epsilon)\right) v_k&=&0\,.\quad\,\,
\end{eqnarray}

\end{appendix}
\bibliography{eta_problem}
\end{document}